\documentclass[conference]{IEEEtran}
\IEEEoverridecommandlockouts
\usepackage{cite}
\usepackage{amsmath,amssymb,amsfonts}
\usepackage{algorithmic}
\usepackage{graphicx}
\usepackage{textcomp}
\usepackage{xcolor}
\usepackage[most]{tcolorbox}
\usepackage{array}
\usepackage{url}
\usepackage{ragged2e}
\newcommand{\sectopic}[1]{\vspace{0.2em}\par\noindent{\textit{\bfseries #1}}}

\def\BibTeX{{\rm B\kern-.05em{\sc i\kern-.025em b}\kern-.08em
    T\kern-.1667em\lower.7ex\hbox{E}\kern-.125emX}}
\begin{document}

\title{Understanding Practitioners' Perspectives on Monitoring Machine Learning Systems\\
\thanks{Identify applicable funding agency here. If none, delete this.}
}

\author{\IEEEauthorblockN{Hira Naveed}
\IEEEauthorblockA{
\textit{Monash University}\\
Australia \\
hira.naveed@monash.edu}
\and
\IEEEauthorblockN{John Grundy}
\IEEEauthorblockA{
\textit{Monash University}\\
Australia \\
john.grundy@monash.edu}
\and
\IEEEauthorblockN{Chetan Arora}
\IEEEauthorblockA{
\textit{Monash University}\\
Australia \\
chetan.arora@monash.edu}
\and

\IEEEauthorblockN{Hourieh Khalajzadeh}
\IEEEauthorblockA{
\textit{Deakin University}\\
Australia \\
hkhalajzadeh@deakin.edu.au}
\and

\IEEEauthorblockN{\centerline{Omar Haggag}}
\IEEEauthorblockA{
\textit{Monash University}\\
Australia\\
omar.haggag@monash.edu}

}

\maketitle

\begin{abstract}

Given the inherent non-deterministic nature of machine learning (ML) systems, their behavior in production environments can lead to unforeseen and potentially dangerous outcomes. For a timely detection of unwanted behavior and to prevent organizations from financial and reputational damage, monitoring these systems is essential. This paper explores the strategies, challenges, and improvement opportunities for monitoring ML systems from the practitioners' perspective. We conducted a global survey of 91 ML practitioners to collect diverse insights into current monitoring practices for ML systems. We aim to complement existing research through our qualitative and quantitative analyses, focusing on prevalent runtime issues, industrial monitoring and mitigation practices, key challenges, and desired enhancements in future monitoring tools. Our findings reveal that practitioners frequently struggle with runtime issues related to declining model performance, exceeding latency, and security violations. While most prefer automated monitoring for its increased efficiency, many still rely on manual approaches due to the complexity or lack of appropriate automation solutions. Practitioners report that the initial setup and configuration of monitoring tools is often complicated and challenging, particularly when integrating with ML systems and setting alert thresholds. Moreover, practitioners find that monitoring adds extra workload, strains resources, and causes alert fatigue. The desired improvements from the practitioners' perspective are: automated generation and deployment of monitors, improved support for performance and fairness monitoring, and recommendations for resolving runtime issues. These insights offer valuable guidance for the future development of ML monitoring tools that are better aligned with practitioners' needs.

\end{abstract}

\begin{IEEEkeywords}
Machine Learning, Monitoring, Industry, Practitioner, Survey
\end{IEEEkeywords}

\section{Introduction}
Machine Learning (ML) systems are being increasingly employed across various domains, including social media, e-commerce, and engineering -- even critical domains such as finance, healthcare, and autonomous vehicles nowadays leverage ML to automate and enhance their services. Generative AI and Large Language Models (LLMs) have further boosted ML adoption by creating several new use cases~\cite{naveed2024model,nguyen2023generative}.       

A typical ML system lifecycle begins by gathering requirements and preparing data, which is followed by the development of the ML component (experimentation, model training, and evaluation) and other traditional software components~\cite{ahmad2023requirements}. After development, the next step is integration and system testing. Once quality assurance is completed, the ML system is deployed to a production environment. This stage is also known as operation or runtime since real users interact with the ML system. The last step of this lifecycle is maintenance, which mainly consists of monitoring and response \cite{shankar2024we}. Monitoring involves continuously observing the behavior of the ML system in production and triggering alerts if any unwanted behavior is detected. Later, response activities are performed to mitigate the runtime issues. The scope of this study is production ML systems and runtime monitoring.

Production ML systems can be particularly difficult to maintain over time due to the brittleness of ML \cite{sculley2015hidden}. Even small changes in the input data or operating environment can significantly impact the outputs of an ML system \cite{sculley2015hidden}. For example, reduced model performance due to changes in the input data distribution. While testing ML systems is valuable, it is insufficient to ensure correct behavior after deployment~\cite{sculley2015hidden}. Monitoring ML systems is essential for early detection and mitigation of runtime issues, thereby maintaining system reliability \cite{sculley2015hidden} and users' trust.
Recent events such as Zillow's house pricing model failure \cite{susarla2024zillow}, Amazon’s biased hiring algorithm \cite{dastin2022amazon}, and Tesla's fatal autopilot accident \cite{bonnefon2016social}, have confirmed the necessity of monitoring ML systems. Considering the financial and reputational losses incurred by large organizations due to runtime issues, ML monitoring is becoming increasingly important in the industry. 

Previous studies have explored monitoring methods and challenges through practitioner interviews and literature reviews \cite{shergadwala2022human,shankar2024we,lewis2021characterizing,heyn2023investigation,schroder2022monitoring,karval2023catching}. These studies reveal challenges such as difficulty customizing monitors, privacy concerns with monitoring, and generating meaningful alerts. However, they do not discuss the complete monitoring context behind these challenges, including the runtime issues encountered, the monitoring and mitigation strategies and tools, associated challenges, and recommended improvements. This gap highlights the need for further research to provide a comprehensive understanding of ML monitoring from an industry perspective. In this study, we present our findings from a survey of 91 ML practitioners about their point of view on ML monitoring. Our survey complements previous research and even validates some of their findings with a broader and more diverse group of participants.

Our findings reveal that practitioners often struggle with runtime issues such as degrading model performance, delayed responses, resource constraints, and security violations. For monitoring approaches, automated solutions were preferred and tools like Prometheus and Grafana were frequently used. Few participants reported to prefer manual monitoring approaches due to the complexity and lack of appropriate automated monitoring solutions, particularly for domain-specific use cases. Additionally, it was mentioned that monitoring adds extra workload on practitioners to learn, integrate, configure, and then manage monitoring tools. Furthermore, monitoring strains system resources, especially storage, due to large volumes of data collected for logging and metrics. We also asked practitioners for improvement recommendations, the responses highlighted better support for performance and fairness monitoring, recommendations for resolving runtime issues, and automated generation and deployment of monitors. 

The main contribution of this study is a holistic overview of ML monitoring based on insights from experienced practitioners across diverse roles, industries, and countries. While previous studies on ML monitoring with human subjects do exist, they were based on interviews with a small number of participants \cite{shergadwala2022human, shankar2024we, heyn2023investigation}. To the best of our knowledge, this is the first study to survey a global cohort of 91 ML practitioners, offering a broader and more current perspective on monitoring practices. The insights from this survey offer valuable guidance for other practitioners and inform the future development of ML monitoring tools that are better aligned with practitioners’ needs.

\sectopic{Structure.} 
The remainder of the paper is structured as follows. Section II summarizes the related work on ML monitoring. Section III describes our research methodology, including research goals, survey design, execution, and data analysis. Section IV reports the findings of our survey and answers the research questions. Section V interprets the survey results and Section VI describes the implications. Section VII explains the threats to the validity of our study and our efforts to mitigate them. Finally, Section VIII concludes the paper.

\section{Related Work}
Prior work has stated that monitoring is critical to MLOps \cite{studer2021towards, kreuzberger2023machine, matsui2022mlops}. Recent studies have explored ML monitoring practices and challenges through surveys, interviews, and literature reviews. 
Zimelewicz et al. \cite{zimelewicz2024ml} perform a global survey with 188 practitioners focusing on ML deployment and monitoring. Regarding monitoring, they found that most production models are not monitored; those that are tend to focus on model inputs and outputs. Key challenges included the lack of established monitoring practices, the need to build custom tools, and difficulty selecting appropriate metrics. These findings are consistent with ours, which offer more detailed insights.

Shergadwala et al. \cite{shergadwala2022human} carried out interviews with 13 ML practitioners to identify the requirements and challenges of monitoring ML models from a human centric perspective. They found that practitioners desire customizations in model monitoring systems for domain-specific use cases, meaningful alerts without cognitive overload, and suggestions for improvement. 
Among challenges, the interviewees mentioned data privacy when using third party monitoring tools, modifying monitoring systems for domain-specific scenarios, and unreliable drift detection. 
In comparison, our study has a broader scope, as it considers runtime issues and monitoring and mitigation strategies in addition to challenges and desired ML monitoring features. Moreover, our survey complements this study by confirming some of the findings and collecting responses from a larger and more diverse pool of participants.

Shankar et al. \cite{shankar2024we} conducted interviews with 18 ML engineers to learn about MLOps tasks and challenges. The study results suggested that ML monitoring and response was the last step in the MLOps workflow. The interviewees described deployment and maintenance of ML models in production as a very iterative, manual, and team-driven task. Considering the manual effort required, three key challenges for ML monitoring and response were highlighted: tracking data quality and investigating alerts, managing complex model pipelines when resolving a production bug, and debugging rare errors. While this study focuses on the entire MLOps lifecycle, our study is specifically about ML monitoring. We investigate beyond challenges to provide a holistic overview of practitioners' experiences with ML monitoring.

Another interview study presented in \cite{heyn2023investigation} explores the challenges of specifying training data and runtime monitors for safety-critical ML systems. The results from the study based on 10 interviews reveal several monitoring challenges. These challenges include a lack of explainability for model decisions, unclear runtime checks, monitoring overhead, and missing technical guidance for ML monitoring in safety standards. 
The scope of this study is restricted to monitoring safety-critical ML systems, whereas our study encompasses monitoring perspectives from various domains.

Lewis et al. \cite{lewis2021characterizing} identified and characterized ML mismatches through an interview of 20 practitioners followed by a validation survey. 
For ML systems in production environments, the study found three mismatches, the first and most prominent one is a lack of runtime metrics, logs, and user feedback for monitoring, reproducing, and correcting runtime errors. The second was unawareness of computing resources available in production, and the third and last one was unawareness of the required time for model inference. Additionally, the authors also explored mismatches related to data in production. 


A systematic literature review in \cite{schroder2022monitoring} reports the challenges and methods of monitoring ML models. They describe monitoring as an extension of testing, as it continuously assesses the system after deployment. From their review, they found that high dimensionality of data during production makes it difficult to ensure quality and thus requires more effort. Shifts in data distribution were another problem since models were deployed under the assumption that production data distribution would be the same as training data. 
Another SLR \cite{karval2023catching} summarized the existing monitoring and explainability methods for ML, and monitoring methods were categorized under data drift, outlier detection, and adversarial detection. Both of these studies provide an overview of monitoring methods, while \cite{schroder2022monitoring} also presents some monitoring challenges.

While existing research discusses the importance of ML monitoring and prevalent challenges, it does not provide a holistic overview of the area. To the best of our knowledge, we are the first to do a global survey to provide a comprehensive understanding of current ML monitoring practices in the industry. By surveying 91 ML practitioners from 11 countries, our study offers a broader and more diverse perspective on monitoring ML systems. We explore various runtime issues in ML systems that necessitate monitoring, monitoring and mitigation strategies, monitoring challenges, and recommendations for improvement.

\section{Methodology}
\subsection{Goal and Research Questions}
The study aims to investigate current industry practices and challenges of monitoring ML systems in production environments. Through a survey of 91 ML practitioners, we collected qualitative and quantitative data to understand practitioners' perspectives regarding various aspects of the monitoring process. This understanding can help other practitioners make informed decisions and guide the future development of ML monitoring tools. The following four research questions (RQs) are addressed in this study:

\textbf{RQ1: What are the common runtime issues practitioners encounter in machine learning systems?} This RQ investigates the runtime problems practitioners experience while working on production ML systems and their causes. This can help create awareness about common runtime issues and the factors contributing towards them, thus encouraging ML monitoring among professionals.

\textbf{RQ2: How do practitioners monitor and mitigate runtime issues in machine learning systems?} This RQ helps understand ML monitoring and issue mitigation strategies prevalent in the industry. We aim to identify and describe the monitoring methods, metrics, tools, and mitigation techniques successfully applied by practitioners.

\textbf{RQ3: What challenges do practitioners face when monitoring machine learning systems?} The insights from this RQ aid in understanding the challenges practitioners encounter when monitoring ML systems. This knowledge is valuable for the improvement of monitoring solutions and can assist other professionals in anticipating and managing similar problems.

\textbf{RQ4: What are the priorities and areas of improvement according to practitioners monitoring machine learning systems? } The findings from this RQ offer insights into practitioners' monitoring preferences and desired improvements. Researchers and organizations developing monitoring solutions can benefit from these insights by better aligning their tools with real-world needs. 

The RQs are conceptually linked to reflect the end-to-end experience of practitioners, from identifying problems and current practices to uncovering pain points and gathering suggestions for improvements.

\subsection{Survey Design and Distribution}
This survey was performed to collect insights from experienced practitioners on ML monitoring practices and challenges. As this research involved human subjects, approval was sought from the university ethics committee before conducting the study.

Our questionnaire consisted of 25 questions, out of which 16 were closed-ended to measure the prevalence of ML monitoring practices, and 9 were open-ended to capture nuanced insights. Quantitative questions were placed first to avoid biasing responses, followed by qualitative questions to allow participants to elaborate on their experiences. The questionnaire was divided into 5 sections, one for each of the following: participant demographics and ML experience (Q1 to Q11), runtime issues encountered (Q12 to Q14), monitoring and mitigation strategies (Q15 to Q21), monitoring challenges (Q22), and monitoring priorities and areas of improvement (Q23 to Q25). Our questionnaire can be found at \cite{ml_monitoring_survey_2025}. We used Qualtrics \cite{qualtrics2025} to design the survey and collect responses.

We conducted a pilot study before launching the survey to assess the feasibility of the survey and refine the questionnaire. It consisted of two ML practitioners recruited through the authors' professional network, whose feedback helped improve the flow and terminology used in the questionnaire. The improved survey was then made publicly available. Responses for the survey were collected in two iterations. During the first iteration, we began by posting adverts on social media websites, LinkedIn, and X. We also searched LinkedIn for professionals with experience of working with ML systems and contacted them by sending a direct message. The messages were significantly more successful in recruiting participants, and we received 63 responses in the first iteration. For more diversity in the responses, we performed a second iteration and recruited 30 participants through Prolific \cite{prolific2025}. The participants of the first iteration did not receive any compensation, however, participants of the second iteration recruited through Prolific were paid £6 per hour. To ensure the quality of responses during the second round, we added 4 additional questions in the survey to check for attention and relevant experience. Two low-quality responses were excluded, resulting in 28 responses from the second round and bringing the total number of responses from both rounds to 91 (63+28).

\subsection{Data Analysis}
We analyzed the data collected from the survey to answer our research questions. For quantitative data from closed-ended questions, we used standard statistical analysis. For qualitative data from open-ended questions, we applied thematic analysis. The first author assigned codes to the responses, grouped similar codes, created themes, and then refined these themes to understand underlying data patterns. All steps were performed under the supervision of other authors, and any disagreements were resolved through consensus.

\section{Results}
\subsection{Participant Demographics and ML Experience}
In this section, we discuss the survey participants' demographic characteristics and ML experience. 

\begin{figure}
    \centering
    \includegraphics[width=1\linewidth]{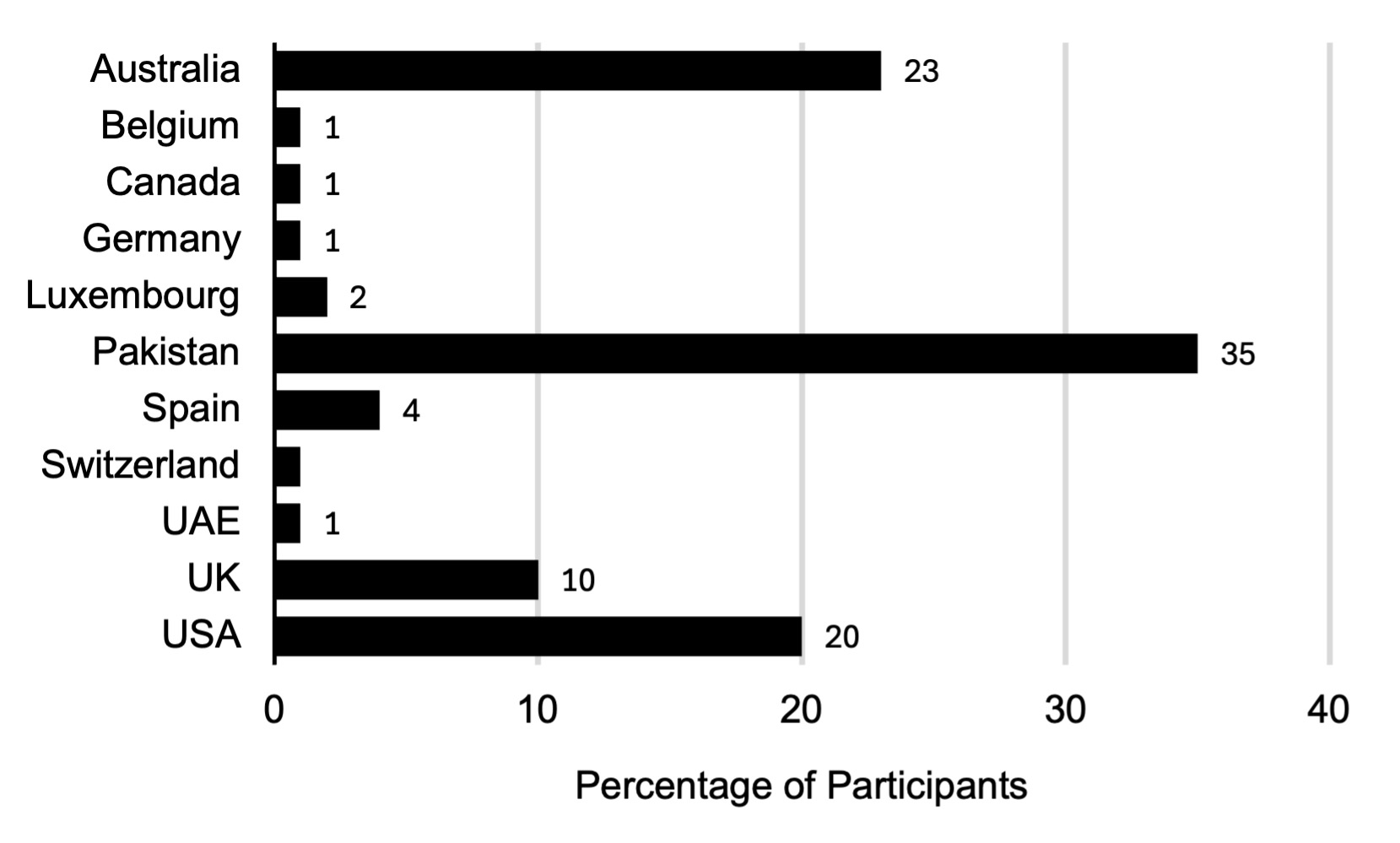}
    \caption{Geographical Location}
    \label{fig:location}
\end{figure}

Figure \ref{fig:location} shows the geographic distribution of participants, with the majority residing in Pakistan (35\%), Australia (23\%), and the United States (20\%). 44\% of the participants were aged 25-34, followed by 25\% aged 18-24, and 19\% aged 35-44. Comparatively, a much smaller portion of respondents were aged 45 and above, 45-54 (4\%), 55-64 (5\%), and 65 or older (2\%). This distribution is consistent with the common age disparity found in software practitioners. In terms of qualification, most of our respondents have a Master's degree (41\%) or a Bachelor's degree (36\%), followed by Doctorate holders (21\%), and a small portion with high school diplomas (2\%). A significant portion of the participants are qualified in Computing and Technology (89\%) fields such as Computer Science, Data Science, Artificial Intelligence, Software Engineering, and Information Technology. The remaining 11\% have qualifications in Engineering or Business.

\begin{figure}
    \centering    \includegraphics[width=1\linewidth]{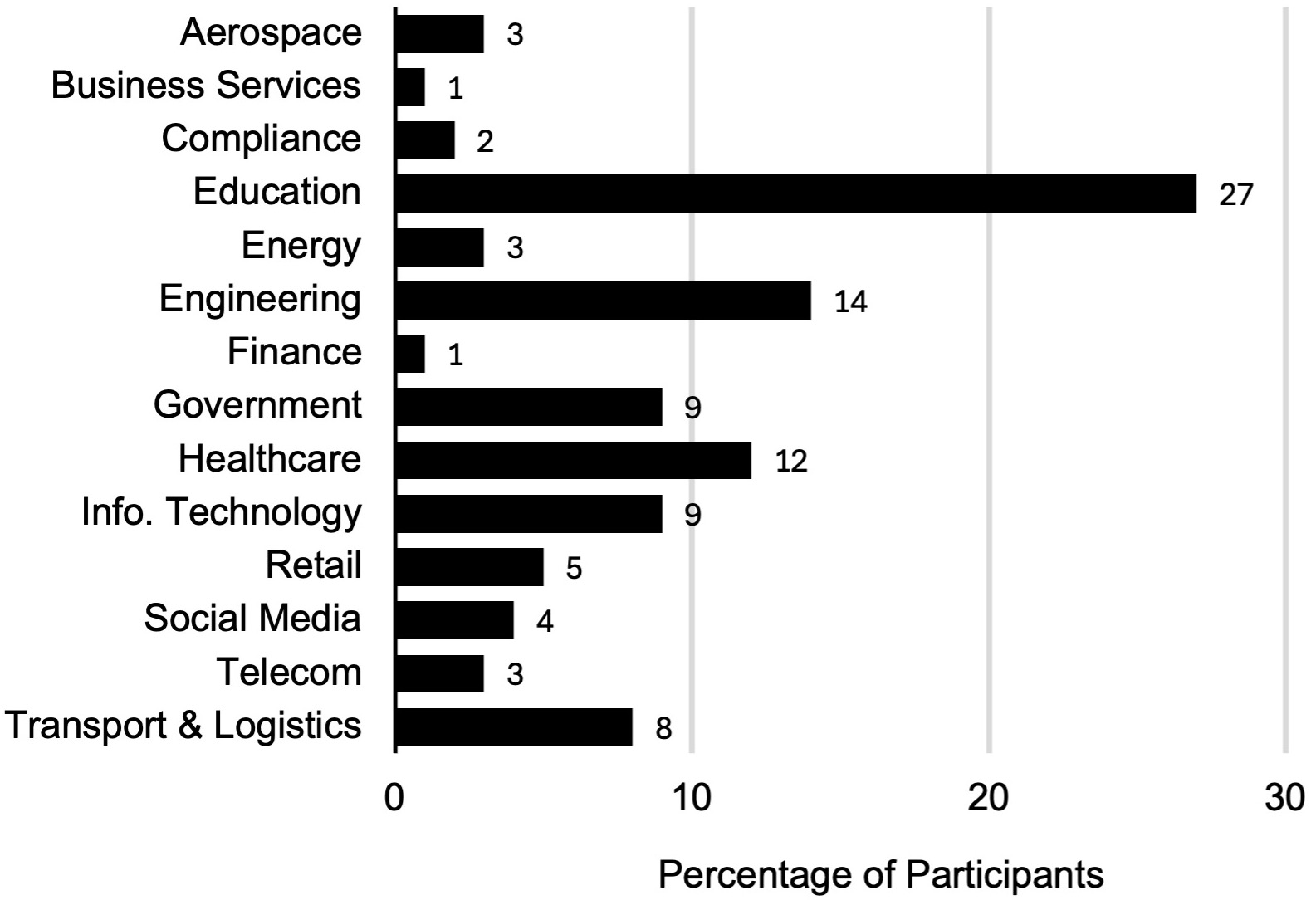}
    \caption{Domain}
    \label{fig:domain}
\end{figure}

Regarding current role, most participants are Machine Learning Engineers (24\%) and Software Developers (23\%). Other designations include Project Manager (14\%), Data Scientist (13\%), DevOps Engineer (11\%), Researcher (7\%), Quality Assurance Engineer (4\%), Business Analyst (1\%), Data Engineer (1\%), IOT Engineer (1\%), Product Owner (1\%), Prompt Engineer (1\%), and Solutions Architect (1\%). The participants have experience across multiple domains, as shown in Figure \ref{fig:domain}. Many respondents reported experience in more than one domain, with the most frequently mentioned ones being Education (27\%), Engineering (14\%), and Healthcare (12\%). 

The practitioners' experience and context of ML systems are important aspects to consider when investigating ML monitoring approaches and challenges, hence, we asked participants about their team size, years of experience, familiarity with ML techniques, goals for leveraging ML, and deployment environments. The results are summarized in Table \ref{tab:experience-demographics}. Considering team size, the largest portion of participants worked in teams of 1-10 people (67\%), followed by those who worked in teams of 11-30 people (22\%). A smaller portion worked in larger teams of 31-50 people (8\%), 51-100 people (2\%), and more than 100 people (2\%). The overall experience of respondents working with ML systems is nearly normally distributed, most have 1-3 years of experience (43\%), or  4-6 years of experience (31\%). A smaller portion are either seasoned experts with 7 or more years of experience (16\%) or in their early career with less than 1 year (10\%). Regarding experience with specific ML techniques, the most prominent ones were Generative AI (22\%) and Supervised Learning (21\%). 

For goals behind employing ML systems, the results indicate an inclination towards using ML for improving products and operational efficiency, with less focus on use cases such as marketing and security. The primary objective reported for leveraging ML systems was Product Development and Innovation (18\%) and Reduce Costs and Optimize Production (18\%), closely followed by Enhance User Experience (16\%) and Quality Improvement and Maintenance (10\%). Comparatively, some less common objectives were Risk Management and Fraud Detection (10\%), Customer Service (9\%), Sales Optimization (8\%), Marketing (5\%), and Security and Compliance (1\%). Regarding deployment environment, participants significantly preferred Cloud-managed Infrastructure (44\%) (e.g., Amazon Web Services (AWS) \cite{aws2024})  over other environments. Similar level of interest was reported for API based ML Services (19\%) (e.g., OpenAI API \cite{openaiapi2024}) and Containerized Deployment (19\%) (e.g., Docker \cite{docker2024} and Kubernetes \cite{kubernetes2024}). Few participants mentioned other deployment environments such as Locally Hosted (12\%) (e.g., on-premise server) and Edge Computing Platform (6\%) (e.g., mobile device).

\begin{table}[ht]
\centering
\caption{Practitioners' Experience and ML Landscape}
\begin{tabular}{p{6cm}r}
\hline
\textbf{\rule{0pt}{8pt}Category} & \textbf{Percentage (\%)} \\
\hline
\multicolumn{2}{l}{\textbf{\rule{0pt}{8pt}Team Size}} \\
1-10 people & 67\% \\
11-30 people & 22\% \\
31-50 people & 8\% \\
51-100 people & 2\% \\
More than 100 people & 1\% \\
\hline
\multicolumn{2}{l}{\textbf{\rule{0pt}{8pt}Experience with ML Systems}} \\
Less than 1 year of experience & 10\% \\
1-3 years of experience& 43\% \\
4-6 years of experience& 31\% \\
7 or more years of experience & 16\% \\

\hline
\multicolumn{2}{l}{\textbf{\rule{0pt}{8pt}Experience with ML Techniques}} \\
Ensemble Learning & 12\% \\
Generative AI & 22\% \\
Reinforcement Learning & 10\% \\
Self-supervised Learning & 8\% \\
Semi-supervised Learning & 13\%\\
Supervised Learning & 21\%\\
Unsupervised Learning & 15\%\\
\hline
\multicolumn{2}{l}{\textbf{\rule{0pt}{8pt}Business Objective for Applying ML}} \\
Enhance User Experience & 16\% \\
Marketing & 5\% \\
Product Development and Innovation & 18\% \\
Quality Improvement and Maintenance & 16\% \\
Reduce Costs and Optimize Production & 18\% \\
Risk Management and Fraud Detection & 10\% \\
Sales Optimization & 8\% \\
Security and Compliance & 1\% \\
\hline
\end{tabular}
\label{tab:experience-demographics}
\end{table}

\subsection{RQ1 - Runtime Issues Encountered}
Practitioners often encounter issues in production ML systems; in this study, we refer to them as \textit{runtime issues}. To better understand these issues, we asked participants whether they had experienced such issues, the specific problems they faced, and what caused them. 70\% of the participants reported that they \textit{have encountered runtime issues}, while 30\% said \textit{they have not}. For those who \textit{have}, we followed up with two open-ended questions about the issues and their causes. Based on the collected data, we identified causes linked with their respective issues and performed thematic analysis to find patterns. Seven themes emerged from this analysis, which complement findings in \cite{zimelewicz2024ml, shergadwala2022human, schroder2022monitoring}.

The first theme, \textbf{Model Performance Issues (38\%)}, demonstrates that practitioners face problems with maintaining the accuracy and correctness of model outputs in production ML systems. This can be due to differences between training and production data (data drift), changes in user patterns or operating context (concept drift), data quality issues such as incorrect format, or unseen edge cases. Issues like data drift were also reported for Generative AI systems, along with hallucinations~\cite{liu2024exploring}. Participant P6 emphasized how reliance on external APIs can be a reason for poor responses, \textit{``With Generative AI solutions, the main reason for this is the dependence on the external LLM APIs, which may be unreliable in certain post-deployment environments."}

The second theme, \textbf{Response Time Issues (25\%)}, suggests delayed responses and limited processing capacity. Key reasons for this include model complexity, infrastructure limitations, and inefficient deployment configurations, often due to an underestimation of production load. For instance, P20 mentioned, \textit{``We identified that if we load an LLM onto separate servers, the latency exceeds tremendously.
 We also identified that if an LLM is loaded on a server where there is already an inference running [...], the latency of the newly loaded LLM gets affected tremendously."}

The third prominent theme, \textbf{Infrastructure Issues (14\%)}, P87 describes it as, \textit{``Infrastructure is unable to handle peak loads (e.g., Black Friday sales crashing recommendation engines)."} This theme emphasizes scalability problems, computational resource constraints, compatibility issues, resource leaks, system crashes, and API cost overruns. Exploring the causes for these issues revealed resource limitations (e.g., CPU/GPU capacity, limited memory), misconfigured cloud infrastructure (e.g., GPU enablement issues), model size and complexity, faulty updates and deployment pipelines, legacy code, and complexity of cloud infrastructure. 

Themes four and five are \textbf{Security and Privacy Violations
 (11\%)} and \textbf{Fairness Violations (8\%)}. The response of P78 highlights the critical nature of such issues, \textit{``Industries that deal with delicate and sensitive data like Healthcare, security violations can have severe repercussions like legal penalties."} Security violations such as data poisoning, data leakage, and model theft are often caused by a lack of security procedures or compliance failures. For fairness violations, they are caused primarily due to insufficient fairness assessment at design time, such as imbalanced training data, biased algorithms, and technical debt~\cite{rezaei2025fairness}.

Theme six \textbf{Consistency and Reliability Issues (3\%)}, refers to inconsistent outputs and unreliable behavior of the ML system that can undermine user trust. P27 pointed out the reason for this, \textit{``Responses from the LLM are not consistent or reliable, which can heavily affect the quality of the application [...] [because] the LLM does not properly comply with the instructions mentioned in the prompt."} 

The last theme, \textbf{Accessibility Issue (1\%)}, illustrates barriers encountered by users who rely on assistive technologies. These occur due to non-compliance with accessibility standards and non-inclusive design choices.

\begin{tcolorbox}[colback=white, colframe=black, boxrule=0.8pt, arc=0pt, left=4pt, right=4pt, top=4pt, bottom=4pt]
\textbf{Answering RQ1:}  The most frequently encountered runtime issues in ML systems, as reported by practitioners, are model output quality issues due to drift, data quality, and unseen edge cases. Response time issues and infrastructure issues are also common concerns, often arising due to resource limitations and higher than expected user volume. While mentioned less often, security, privacy, and fairness violations remain an important concern, particularly in critical applications. A few respondents also highlighted consistency, reliability, and accessibility issues.
\end{tcolorbox}

\begin{table*}[h]
\renewcommand{\arraystretch}{1.7}
\caption{Monitoring and Mitigation Approaches}
\label{tab:monitoring-approaches}
\centering
\footnotesize
\begin{tabular}{|>{\raggedright\arraybackslash}p{2.2cm}|>{\raggedright\arraybackslash}p{2.8cm}|>{\raggedright\arraybackslash}p{2.5cm}|>{\raggedright\arraybackslash}p{2.5cm}|>{\raggedright\arraybackslash}p{2.8cm}|>{\raggedright\arraybackslash}p{3.2cm}|}

\hline
\textbf{Category} & \textbf{Specific Requirements} & \textbf{Techniques \& Metrics} & \textbf{Primary Tools}  & \textbf{Data Collected} & \textbf{Mitigation Strategies} \\
\hline

\textbf{Model Performance} & Model accuracy, precision, recall, F1-score, AUC-ROC, RMSE, prediction quality & Statistical metrics, confusion matrices, error rate analysis, threshold-based monitoring & MLflow \cite{mlflow}, Weights \& Biases \cite{wandb}, Prometheus \cite{prometheus}, Grafana \cite{grafana} & Model predictions and ground truth, model I/O pairs, confidence scores & Model retraining, hyperparameter tuning, automated pipelines, rollback mechanisms \\
\hline
\textbf{Data Quality \& Drift} & Data drift, feature drift, data format & Statistical drift tests (Kolmogorov Smirnov test \cite{berger2014kolmogorov}, Jensen-Shannon test \cite{menendez1997jensen}), distribution comparisons, schema validation & Evidently AI \cite{evidently}, MLflow \cite{mlflow}, Alibi Detect \cite{alibidetect}, custom scripts & Feature distributions, data statistics, correlations, missing values & Automated retraining triggers, pipeline adjustments, feature engineering, quality checks \\
\hline
\textbf{Response Time} & End-to-end latency, inference time, API response time, processing speed & P95/P99 latency percentiles, threshold-based monitoring & Prometheus \cite{prometheus}, Grafana \cite{grafana}, AWS CloudWatch \cite{aws2024}, Azure App Insights \cite{azureappinsights}, DataDog \cite{datadog} & Response times, API latency logs, inference duration, performance logs & Infrastructure scaling (Kubernetes \cite{kubernetes2024}), model optimization (TensorRT \cite{tensorrt}, ONNX \cite{onnx}), caching (Redis \cite{redis}), load balancing \\
\hline
\textbf{Resource Utilization} & CPU/GPU usage, memory consumption, disk usage, throughput, system load & Resource monitoring, utilization percentages & Prometheus \cite{prometheus}, Grafana \cite{grafana}, AWS CloudWatch \cite{aws2024}, Azure Monitor \cite{azureappinsights}, DataDog \cite{datadog} & CPU/GPU/memory stats, disk I/O, network utilization, performance data & Auto-scaling, resource optimization, load balancing, containerization (Docker \cite{docker2024}/Kubernetes \cite{kubernetes2024}) \\
\hline
\textbf{System Reliability} & Uptime, availability, error rates, system crashes, service failures & Uptime monitoring, error tracking, health checks, availability metrics & Pingdom \cite{pingdom}, AWS CloudWatch \cite{aws2024}& System logs, error traces, uptime statistics, health results, failure logs & Redundancy, failover mechanisms, circuit breakers, automated recovery \\
\hline
\textbf{Security \& Privacy} & Adversarial attacks, privacy violations, prompt injections, data leakage & Anomaly detection, security attacks, privacy assessments & AWS WAF \cite{awswaf}, security scanners, Security information and event management (SIEM) & Security logs, anomalous patterns, breach reports, access logs & Security patches, input filtering, rate limiting, API key rotation, access controls \\
\hline
\textbf{Fairness \& Bias} & Demographic parity, disparate impact, bias across groups, equitable outcomes & Statistical testing, Fairness metrics, bias audits, group-based analysis & Fairlearn \cite{fairlearn}, AI Fairness 360 \cite{aif360},  & Demographic data, prediction outcomes by group, bias indicators, fairness scores & Bias mitigation algorithms, data rebalancing, fairness-aware training, diverse training data \\
\hline
\textbf{Cost } & Token usage, API costs, computational expenses, resource efficiency & Cost tracking, usage monitoring, efficiency metrics & OpenAI dashboard \cite{openaidashboard}, billing APIs \cite{openaiapi2024} & Token usage, API logs, billing data, consumption costs, usage patterns & Prompt optimization, caching, rate limiting, resource optimization, cost alerts \\
\hline
\end{tabular}
\end{table*}

\subsection{RQ2 - Monitoring and Mitigation Strategies}
\begin{figure}
    \centering
    \includegraphics[width=1\linewidth]{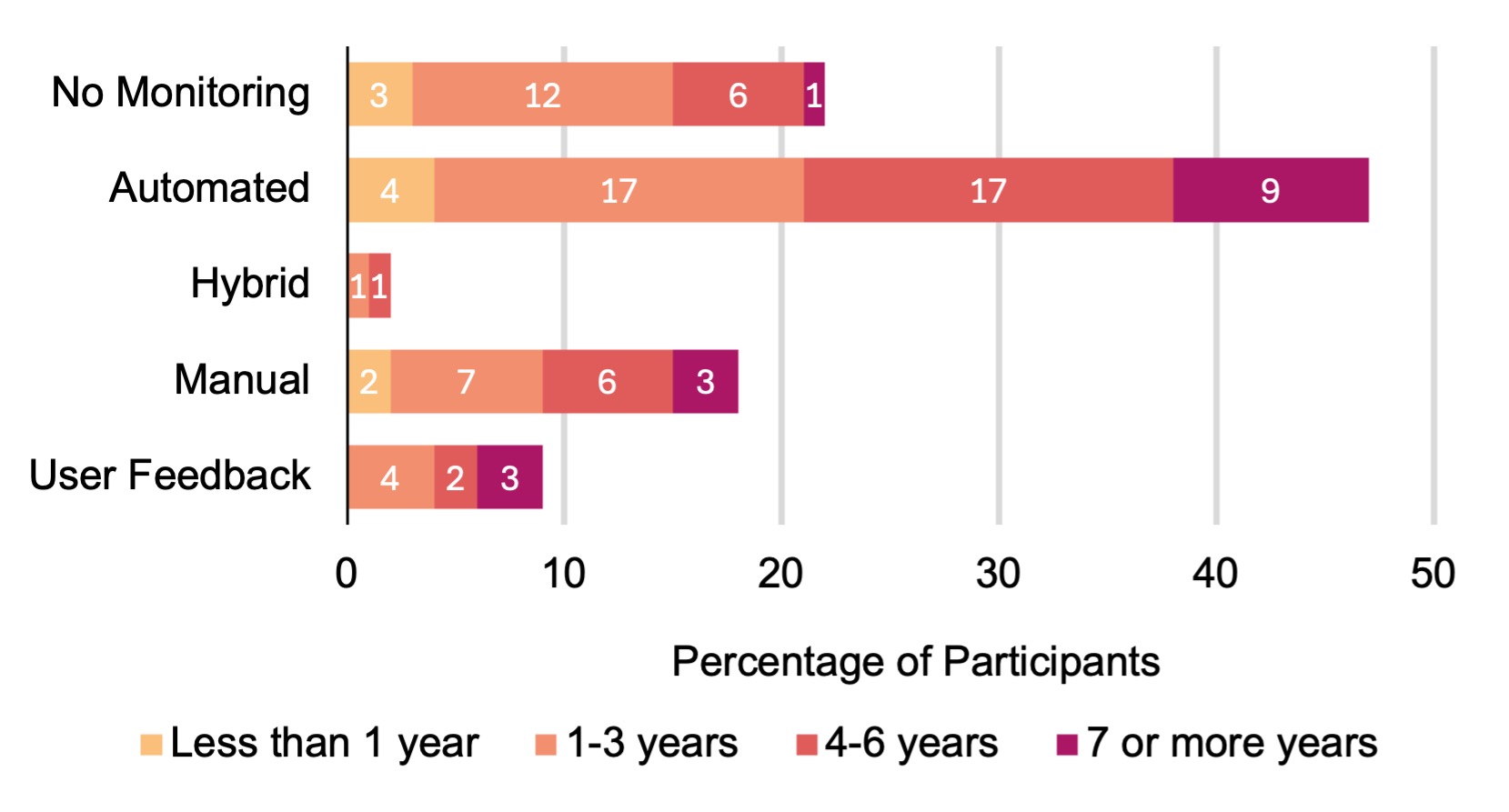}
    \caption{Monitoring Technique with respect to Experience Level}
    \label{fig:technique}
\end{figure}

To understand how practitioners monitor ML systems, we asked them two closed-ended and three open-ended questions. The first question was about whether they monitor ML systems or not. 77\% of participants reported that they \textit{do monitor production ML systems}, while 23\% said they \textit{do not}. Comparing these responses with participants' experiences of runtime issues, we found that 58\% of those who monitor ML systems have also encountered runtime issues, while 19\% monitor but have not experienced such issues. A smaller portion of 10\% participants reported encountering runtime issues despite not monitoring their systems, and 11\% neither monitor nor report having faced runtime issues. For those who \textit{did monitor}, we asked a follow-up question about the monitoring techniques used in practice. Figure \ref{fig:technique} shows the results of various techniques applied while also considering the experience level of participants. Automated monitoring techniques (47\%) were preferred by the majority of participants, followed by manual monitoring (19\%), and reliance on user feedback (9\%). A small group reported using Hybrid techniques (2\%) that combine automated, manual, and user feedback-based monitoring. Interestingly, the experience level of practitioners did not appear to influence their choice of monitoring technique; some junior practitioners used automated methods, while some experienced practitioners continued to rely on manual approaches.

Table \ref{tab:monitoring-approaches} presents a summary of our findings on ML monitoring strategies and approaches to mitigate runtime issues. The results are categorized according to the monitored aspect and divided into eight categories. One notable trend is the heavy reliance on statistical metrics and threshold-based monitoring for model performance and drift monitoring. We observed that basic threshold-based monitoring was also frequently used to detect latency issues. Open-source tools such as MLflow \cite{mlflow}, Prometheus \cite{prometheus}, and Grafana \cite{grafana} are widely used to track model performance and resource metrics. Data drift detection is typically done through data distribution comparisons, and response times are monitored through latency profiles. These monitoring strategies often trigger automated mitigation actions like model retraining, scaling, and caching.

Another trend is the growing awareness of responsible AI aspects such as fairness, privacy, and cost monitoring. Although these areas are monitored far less frequently than model performance, drift, or latency, they are gaining attention. Monitoring these aspects typically relies on specialized tools, such as Fairlearn \cite{fairlearn}, AI Fairness 360 \cite{aif360}, and AWS WAF \cite{awswaf}. Cost and resource monitoring are often supported through usage tracking dashboards like the OpenAI dashboard \cite{openaidashboard} and cloud billing APIs, with mitigation strategies focused on prompt optimization and resource efficiency. Overall, the ML monitoring landscape reflects a balance between model performance and system operation concerns, with growing interest in responsible ML aspects like fairness.

\begin{figure}
    \centering
    \includegraphics[width=1\linewidth]{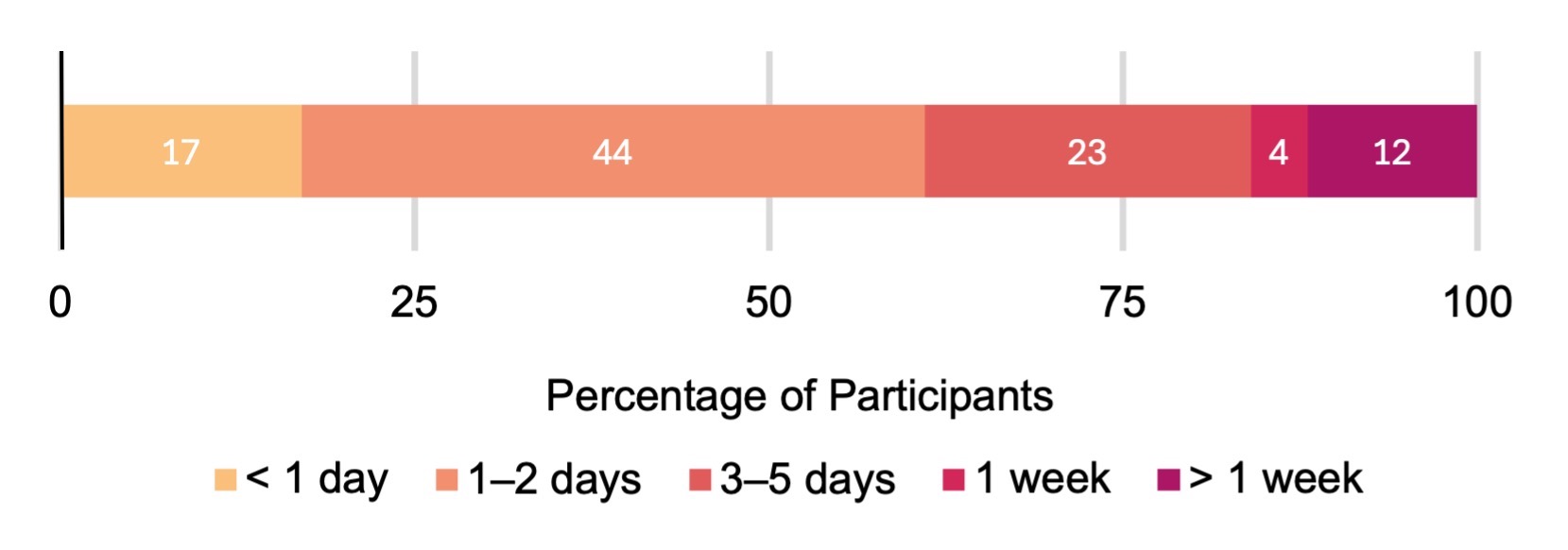}
    \caption{Time Taken To Identify and Mitigate Runtime Issues}
    \label{fig:mitigation}
\end{figure}

Lastly, we asked participants how long it takes for them to identify and mitigate runtime issues. The responses summarized in Figure \ref{fig:mitigation}, show that 44\% of participants reported it takes them between 1-2 days for the complete detection and mitigation process. 23\% shared it takes them between 3 to 5 days, 16\% said less than a day, 12\% reported more than a week, and 4\% stated exactly one week. Overall, for 40\% of participants, identifying and resolving runtime issues takes between three days to over a week. This is a prolonged time frame that can lead to financial losses and reduced user trust in the ML systems.

\begin{tcolorbox}[colback=white, colframe=black, boxrule=0.8pt, arc=0pt, left=4pt, right=4pt, top=4pt, bottom=4pt]
\textbf{Answering RQ2:} Most practitioners preferred automated monitoring techniques, though some still relied on manual methods and user feedback. Interestingly, no relationship was found between practitioners' experience level and choice of monitoring technique. The primary monitoring categories for practitioners include model performance, response time, resource utilization, and drift. 
Monitoring tools like Prometheus and Grafana were most commonly reported for performance, response time, and resource utilization monitoring. MLflow was also a popular choice for monitoring performance and drift. 
Among mitigation strategies, model retraining was found to be useful for addressing performance and drift issues; scaling and optimization were used for infrastructure and resource issues; and redundancy and automated recovery helped with reliability issues. 
For nearly half of the participants, the complete process of identifying and resolving runtime issues took between three days to over a week.
\end{tcolorbox}
\subsection{RQ3 - Monitoring Challenges } \label{challenges}
Monitoring production ML systems presents several challenges for practitioners \cite{shankar2024we, shergadwala2022human}. We asked participants who monitor ML systems about these challenges through an open-ended question. Out of all the participants, 34\% responded to this question, and their responses revealed four key themes:

The first theme, \textbf{Operational Overhead (40\%)}, represents the additional effort and cost involved in monitoring. P3 noted, \textit{``Automated monitoring for such [Retrieval Augmented Generation (RAG)] systems is limited as the answers vary from domain to domain, therefore generic monitors are not feasible."} Participants mentioned increased workload when learning new monitoring tools, manually modifying them, and dealing with alert fatigue due to false alarms. P21 shared, \textit{``[...] Using it [monitoring solution] was difficult as it had very limited functionality and required significant time to implement new monitoring requirements."} Additional concerns included the cost of monitoring infrastructure, high computational requirements, data labeling costs, resource constraints, and scalability challenges in storing and analyzing large volumes of monitoring data, also observed in~\cite{heyn2023investigation,shankar2024we}.

The second theme, \textbf{Setup and Configuration Complexity (31\%)}, captures the challenges practitioners face during the initial setup and configuration of monitors. Similar to \cite{zimelewicz2024ml, shergadwala2022human}, we found that this process can be manual and tedious, especially selecting appropriate metrics, identifying which properties to monitor, configuring alert thresholds, collaborating with other team members, setting up automated alerts, and balancing monitoring tradeoffs. For example, P90 mentioned, \textit{``[...] Ensuring scalable, accurate monitoring of diverse queries on platforms like grok.com is particularly tricky."} Similarly, P84 reported, \textit{``[...] setting up accurate drift detection, dealing with noisy alerts, and ensuring the monitoring dashboards were actually useful without being overwhelming were definitely challenges."}

The third theme, \textbf{Integration Difficulties (21\%)}, represents the challenges in integrating monitoring tools with the ML system, particularly, automating data streams to monitors. For instance, P83 pointed out, \textit{``Integration [of monitoring tools] with existing CI/CD pipelines was difficult. [There is a] lack of plug-and-play monitoring tools for small-scale systems [...]"}.

Lastly, the fourth theme, \textbf{Data Quality and Availability Issues (8\%)}, encompasses challenges such as the unavailability of ground truth data, inconsistencies in data quality, and privacy concerns during data logging and analysis, also reported in \cite{zimelewicz2024ml, shergadwala2022human}. For example, P47 reported, \textit{``Ensuring fairness across [...] subgroups was difficult due to the lack of balanced ground truth data for all categories."}

\begin{tcolorbox}[colback=white, colframe=black, boxrule=0.8pt, arc=0pt, left=4pt, right=4pt, top=4pt, bottom=4pt] 
\textbf{Answering RQ3:} Two major pain points in ML monitoring, as reported by practitioners, are the operational overhead and setup and configuration complexity. Participants described these aspects as laborious, time-consuming, and contributing to increased workload, particularly due to false alarms. Integration of monitoring tools with existing ML systems was also mentioned as a challenge by a smaller portion of respondents. In comparison, issues related to data quality and the unavailability of labelled data were not a major concern.  
\end{tcolorbox}

\subsection{RQ4 - Monitoring Priorities and Areas of Improvement}
To improve the ML monitoring experience for practitioners, it is essential to understand their preferences and requirements. We asked participants two closed-ended questions, each including an \textit{Other} option to allow for additional input.

\begin{figure}
    \centering
    \includegraphics[width=1\linewidth]{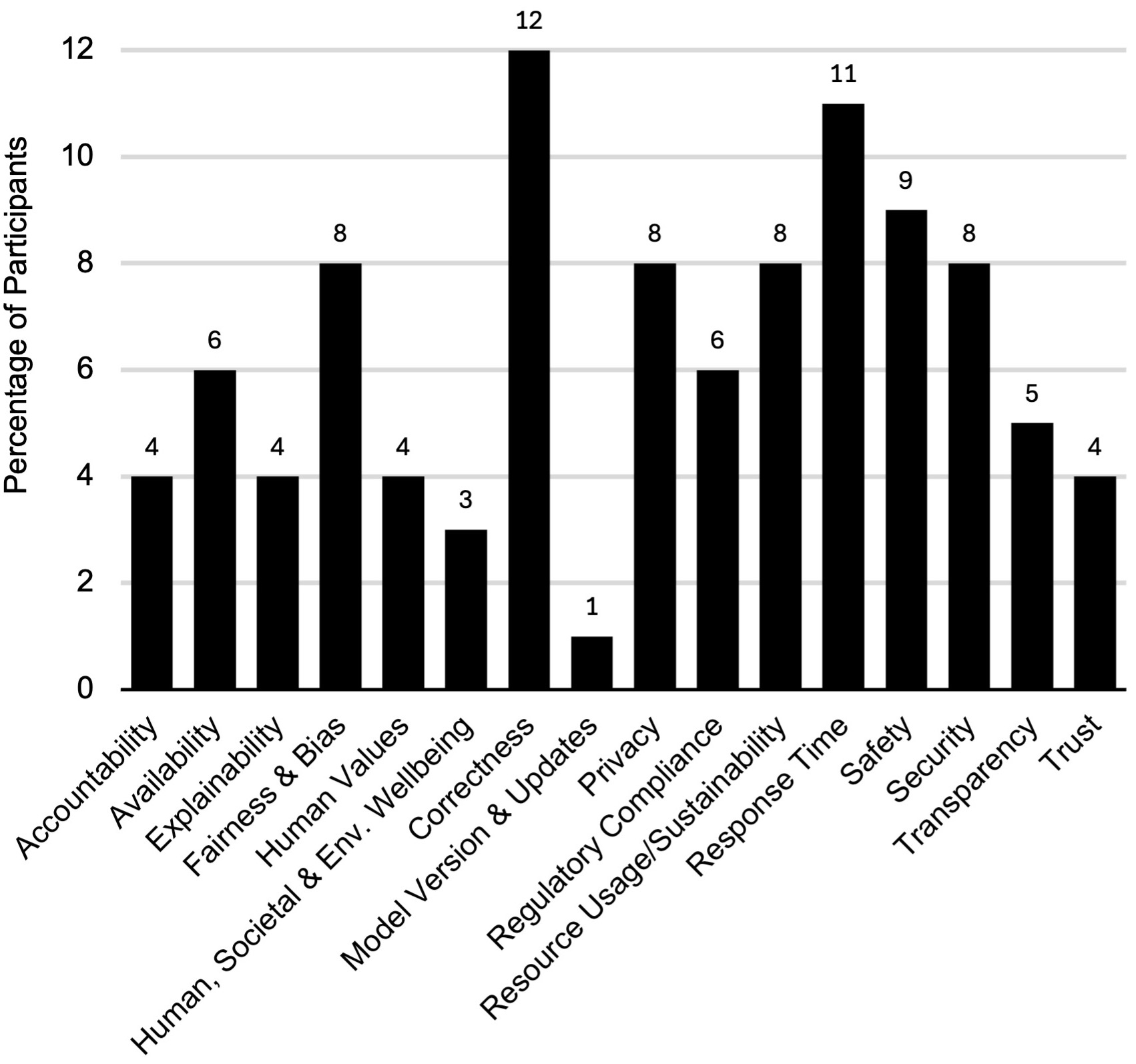}
    \caption{Monitoring Priorities}
    \label{fig:desired}
\end{figure}

Figure \ref{fig:desired} shows the monitoring aspects practitioners prioritize in ML systems. Model Performance and Response Time continue to dominate with 12\% and 11\% respectively. Participants also showed a substantial and nearly similar interest in Safety (9\%), Fairness and Bias (8\%), Privacy (8\%), Resource Usage/Sustainability (8\%), and Security (8\%). Aspects that received decent attention were Availability (6\%), Regulatory Compliance (6\%), Transparency (5\%), Accountability (4\%), Explainability (4\%), Human Values (4\%), Trust (4\%), and Human Values (3\%). This trend shows a growing awareness regarding responsible ML among practitioners. Tracking Model Version and Updates (1\%) was cited using the \textit{Other} option by only one participant.
\begin{table}[ht]
\centering
\caption{Areas of Improvement}
\begin{tabular}{p{6cm}r}
\hline
\textbf{\rule{0pt}{8pt}Support Area} & \textbf{Percentage (\%)} \\
\hline
\rule{0pt}{8pt}Automated Monitor Setup \& Deployment & 19\% \\
Automated Resolution of Runtime Issues & 10\% \\
Explanations for Runtime Issues \& Root Causes & 7\% \\
Fairness \& Bias Monitoring & 13\% \\
Model Performance Monitoring & 18\% \\
Privacy Monitoring & 9\% \\
Recommended Fixes for Runtime Issues & 12\% \\
Traceability across Business Objectives, Requirements, \& Metrics & 12\% \\
\hline
\end{tabular}
\label{tab:monitoring-support-areas}
\end{table}

Table \ref{tab:monitoring-support-areas} presents the areas of improvement for ML monitoring solutions. The findings indicate a strong interest in Automated Monitor Setup and Deployment (19\%), this insight is consistent with the challenges reported regarding the complexity of configuring monitors. Model Performance Monitoring (18\%) is the second most mentioned area of improvement according to practitioners. This aligns with the frequent runtime issues practitioners experience with model performance. Other popular areas with the potential for improvement are Fairness and Bias Monitoring (13\%), Recommended Fixes for Runtime Issues (12\%), and Traceability across Business Objectives, Requirements, and Metrics (12\%). Less common areas are Automated Resolution of Runtime Issues (10\%), Privacy Monitoring (9\%), and Explanations for Runtime Issues and Root Causes (7\%). Few participants used the \textit{Other} option to recommend additional features. 
P58 shared, \textit{``Intuitive interface for setting up and managing monitors to reduce technical barrier for team members, a robust system for automated fixes and recommendations would help streamline operations and improve system reliability [...]. Understanding the root causes of requirements violations in clearer terms would enable us to make informed adjustments and enhance overall system performance."} Similarly, P33 noted, \textit{``Efficient real-time monitoring, automated remediation, and a focus on transparency and explainability are key to addressing these [runtime] issues effectively."}

\begin{tcolorbox}[colback=white, colframe=black, boxrule=0.8pt, arc=0pt, left=4pt, right=4pt, top=4pt, bottom=4pt] 
\textbf{Answering RQ4:} Practitioners largely prioritize monitoring model performance and response time over other aspects. There is growing interest among practitioners to monitor responsible ML aspects such as safety, fairness and bias, privacy, sustainability, and transparency. Qualities such as security remain important as well, while accountability, explainability, and human values received less attention. Regarding areas of improvement, automated monitor generation and deployment, and better performance monitoring were frequently cited. In contrast, fewer participants wanted improved fairness monitoring, recommendations of fixes, and traceability between requirements and metrics. 
\end{tcolorbox}

\section{Discussion}
\subsection{Low-Code Monitor Setup and Management}
Our findings revealed that practitioners struggle with the initial setup and configuration of monitors. The increased workload of learning new tools, dealing with false alarms, and manually tuning thresholds highlights a need for automation. Practitioners expressed a strong desire for automated solutions that would reduce technical barriers and time to set up monitors. Low-code \cite{prinz2021low} or no-code \cite{moskal2021no} approaches would be ideal in this regard for quick monitor creation, deployment, and management \cite{naveed2024model}. While few approaches exist \cite{kourouklidis2020towards}, they do not sufficiently address all aspects of this issue described in the challenges section \ref{challenges}. 
\subsection{Generative AI Specific Monitoring Solutions}  The unique nature of generative AI systems requires monitoring solutions tailored to these systems. Participants mentioned that they often monitor generative AI systems manually because it is difficult to find appropriate automated tools. Since generative AI is a relatively new area, available monitoring solutions and tools are limited \cite{datadog}, particularly open-source options \cite{deepeval}. There is a clear and urgent need for more targeted monitoring solutions focusing on generative AI systems. These solutions should cater to their unique characteristics, such as handling natural language prompts and responses, efficiently managing high computational loads, and minimizing token usage costs if using approaches like LLM-as-a-Judge \cite{gu2024survey}.

\subsection{Domain-specific Monitoring Solutions}
Based on our findings, it is evident that domain-specific monitoring solutions are scarce. Participants reported that they either rely on manual monitoring or have to write scripts themselves for monitoring domain-specific ML applications. Additionally, configuring thresholds can be particularly challenging for such use cases, causing alert fatigue among practitioners. This emphasizes the need for monitoring solutions that can be easily adapted for domain-specific scenarios and provide threshold recommendations while balancing tradeoffs.

\subsection{Disconnect between Monitoring Solutions and Adoption} Our findings revealed that practitioners prioritize model performance above all, and it is also the area where most runtime issues are encountered. While several approaches and tools for model performance monitoring exist \cite{eck2022monitoring, sopan2021ai}, practitioners continue to face challenges even with well-known and well-researched issues like data drift \cite{eck2022monitoring} and concept drift \cite{sopan2021ai}. Our study did not collect sufficient data to determine whether these challenges are due to a lack of awareness, maturity of solutions, cost barriers, or any other factors. Regardless of the cause, this highlights a major disconnect between research and practice. Future works should explore the underlying reasons for this gap and focus on bridging it.

\subsection{Increased Awareness of Responsible ML but Limited Adoption} 
Monitoring priorities of practitioners suggested an increased awareness of responsible ML monitoring. While it is encouraging to see this awareness, the implementation in practice was limited. Only a small portion of participants reported monitoring for responsible ML aspects such as fairness and accessibility. One possible reason is the additional effort required to implement and manage these monitors. Integrating features for responsible ML into existing monitoring tools could reduce this burden, eliminating the need for separate systems. Additionally, developing user-friendly, plug-and-play–style solutions for responsible ML monitoring would help encourage broader adoption.

\subsection{Design with Monitoring in Mind}
Integration of monitoring tools with existing ML systems is one of the challenges reported by practitioners. A possible reason for this is that many ML system architectures were not originally designed with future monitoring requirements in mind. Proactively accounting for monitoring needs during the design of ML systems and CI/CD pipeline would greatly simplify integration in later stages. 

\section{Implications}
To enhance the practical relevance of our findings, we outline key implications below.
\subsection{Practitioners}
Our study provides practitioners with an overview of common runtime issues encountered in production ML systems. Practitioners should prioritize these issues based on severity according to their specific context and identify monitoring requirements early in the ML development lifecycle. Early consideration would allow ML systems to be developed in a way that supports built-in instrumentation or hooks aligned with suitable monitoring tools, enabling smoother integration and reducing operational efforts or the need for major changes later. Table \ref{tab:monitoring-approaches} provides a starting point by summarizing various monitoring aspects, techniques, tools, and mitigation strategies used by other practitioners.
\subsection{Tool Builders}
Our findings suggest that tool developers should prioritize low-code and domain-adaptable monitoring solutions that reduce operational overhead, simplify integration, and offer built-in support for responsible ML aspects such as fairness and privacy. Automation in monitor setup, accurate alerts, issue mitigation suggestions, and support for newer system types (e.g., generative AI) are also critical design needs. 
\subsection{Researchers}
For researchers, the identified monitoring challenges and improvement areas provide a foundation for evaluating existing tools and developing frameworks aligned with practitioners' needs. It would also be valuable to investigate why model performance monitoring challenges persist despite the availability of several tools, and how to bridge the gap between tool capabilities and practitioner needs.

\section{Threats to Validity}
In this section, we discuss the threats to the validity of this study and our attempts to mitigate them.

\subsection{Internal Validity}
A potential threat to internal validity is that factors such as participants' experience level, team size, domain, or demographic location of the practitioners may influence the results. To reduce this, we collected responses from 11 countries, with the top 4 countries representing different global regions. To prevent the impact of participants' experience, we collected responses from novices, mid-level, and experts in the field of ML. Our participant pool also covers a range of application domains and team sizes for better representation. However, it is important to note that these factors were self-reported, and while the responses are diverse across multiple categories, the distribution is not uniform, which may affect the results. A specific risk associated with participants recruited through Prolific was the potential to misrepresent their experience or rush through the survey. To address this, we added additional attention and relevant experience check questions and excluded responses that did not correctly answer these questions. 

\subsection{External Validity}
Threats to external validity include results not being generalizable. Since this was an anonymous survey, we could not collect organizational affiliations of participants to guarantee the authenticity of industrial experience. However, 
many participants were recruited through LinkedIn, where professional affiliations are publicly visible. While this does not entirely eliminate the risk, we have tried to reduce its impact on the study. Additionally, recruitment through social media and anonymous surveys is a commonly accepted practice in software engineering research involving industry participants~\cite{nogueira2024insights, peruma2024developer}. We tried to include participants from a range of domains, team sizes, experience levels, and geographical locations. To reach a broader audience, we recruited participants through social media and Prolific. However, 85\% of our participants are from Pakistan, Australia, and the United States, which is a threat to the generalizability of the study, as regional differences may influence the reported results. Another concern was selection bias, as the practitioners who responded might be highly experienced in ML monitoring. To mitigate this, we recruited participants based on ML experience, not ML monitoring experience. Lastly, for RQ3, since only 34\% of participants responded, the results may not fully represent all monitoring challenges that exist. To mitigate this, we analyzed responses from diverse roles, experience levels, and domains to reflect varied practitioner perspectives.

\subsection{Construct Validity}
The terminology used in survey questions is a potential threat to construct validity. We addressed this in three ways: (i) we added a detailed explanatory statement with examples at the beginning of the survey; (ii) we conducted a pilot study with experienced ML practitioners to ensure that the survey terminology was consistent with industry terminology; and (iii) during analysis of free-text answers, if the authors observed that a participant had not correctly understood the question, their response was excluded from the survey. Additionally, closed-ended questions can also be a concern for construct validity, as they limit the range of answers. To lower this risk, we added a free-text ``other" option with nearly all such questions. The same questions and ordering were used for both survey iterations, with the addition of 4 questions (attention and relevant experience checks) for Prolific. 

\subsection{Conclusion Validity}
A major threat to conclusion validity is false or inaccurate reporting by participants, especially those recruited through Prolific. For issues of genuine misunderstanding, we performed pilot tests to refine survey terminology and ensure it was easily understood by practitioners. While this risk cannot be completely mitigated, we have tried to exclude misreported and misunderstood responses where possible. Lastly, our sample size offers reasonable coverage for capturing diverse practitioner perspectives.

\section{Conclusion and Future Work}
This paper investigates the complete monitoring landscape for production ML systems. This includes runtime issues, monitoring approaches and tools, mitigation strategies, monitoring challenges, priorities, and areas of improvement. Our goal was to understand how ML practitioners perform monitoring and how it can be improved. By conducting a survey with 91 practitioners, we collected diverse insights to identify current industrial practices, pain points, and opportunities for advancement. 
The findings from this survey are beneficial for other practitioners and can help guide the design of future ML monitoring tools that more closely reflect real-world needs. 

Potential future work directions include follow-up interviews for deeper analysis into the types of ML systems used, runtime issues specific to each type of system, and why certain monitoring tools are preferred.
Further exploration of monitoring for generative AI and responsible ML would also be valuable.

\section*{Acknowledgment}
Naveed is supported by a Faculty of IT Post-graduate scholarship.
Grundy and Haggag are supported by ARC Laureate Fellowship
FL190100035. Haggag is also supported by a National Intelligence Post-doctoral Fellowship. This work is also partly supported by ARC Discovery
Project DP200100020.

\bibliographystyle{IEEEtran} 
\bibliography{references}

\begin{thebibliography}{10}
\providecommand{\url}[1]{#1}
\csname url@samestyle\endcsname
\providecommand{\newblock}{\relax}
\providecommand{\bibinfo}[2]{#2}
\providecommand{\BIBentrySTDinterwordspacing}{\spaceskip=0pt\relax}
\providecommand{\BIBentryALTinterwordstretchfactor}{4}
\providecommand{\BIBentryALTinterwordspacing}{\spaceskip=\fontdimen2\font plus
\BIBentryALTinterwordstretchfactor\fontdimen3\font minus \fontdimen4\font\relax}
\providecommand{\BIBforeignlanguage}[2]{{%
\expandafter\ifx\csname l@#1\endcsname\relax
\typeout{** WARNING: IEEEtran.bst: No hyphenation pattern has been}%
\typeout{** loaded for the language `#1'. Using the pattern for}%
\typeout{** the default language instead.}%
\else
\language=\csname l@#1\endcsname
\fi
#2}}
\providecommand{\BIBdecl}{\relax}
\BIBdecl

\bibitem{naveed2024model}
H.~Naveed, C.~Arora, H.~Khalajzadeh, J.~Grundy, and O.~Haggag, ``Model driven engineering for machine learning components: A systematic literature review,'' \emph{Information and Software Technology}, vol. 169, 2024.

\bibitem{nguyen2023generative}
A.~Nguyen-Duc, B.~Cabrero-Daniel, A.~Przybylek, C.~Arora, D.~Khanna, T.~Herda, U.~Rafiq, J.~Melegati, E.~Guerra, K.-K. Kemell \emph{et~al.}, ``Generative artificial intelligence for software engineering--a research agenda,'' \emph{arXiv preprint arXiv:2310.18648}, 2023.

\bibitem{ahmad2023requirements}
K.~Ahmad, M.~Abdelrazek, C.~Arora, M.~Bano, and J.~Grundy, ``Requirements practices and gaps when engineering human-centered artificial intelligence systems,'' \emph{Applied Soft Computing}, vol. 143, 2023.

\bibitem{shankar2024we}
S.~Shankar, R.~Garcia, J.~M. Hellerstein, and A.~G. Parameswaran, ``" we have no idea how models will behave in production until production": How engineers operationalize machine learning,'' \emph{Proceedings of the ACM on Human-Computer Interaction}, vol.~8, no. CSCW1, 2024.

\bibitem{sculley2015hidden}
D.~Sculley, G.~Holt, D.~Golovin, E.~Davydov, T.~Phillips, D.~Ebner, V.~Chaudhary, M.~Young, J.-F. Crespo, and D.~Dennison, ``Hidden technical debt in machine learning systems,'' \emph{Advances in neural information processing systems}, vol.~28, 2015.

\bibitem{susarla2024zillow}
P.~Susarla, D.~Purnell, and K.~Scott, ``Zillow’s artificial intelligence failure and its impact on perceived trust in information systems,'' \emph{Journal of Information Technology Teaching Cases}, 2024.

\bibitem{dastin2022amazon}
J.~Dastin, ``Amazon scraps secret ai recruiting tool that showed bias against women,'' in \emph{Ethics of data and analytics}.\hskip 1em plus 0.5em minus 0.4em\relax Auerbach Publications, 2022.

\bibitem{bonnefon2016social}
J.-F. Bonnefon, A.~Shariff, and I.~Rahwan, ``The social dilemma of autonomous vehicles,'' \emph{Science}, vol. 352, no. 6293, 2016.

\bibitem{shergadwala2022human}
M.~N. Shergadwala, H.~Lakkaraju, and K.~Kenthapadi, ``A human-centric perspective on model monitoring,'' in \emph{Proceedings of the AAAI Conference on Human Computation and Crowdsourcing}, vol.~10, 2022.

\bibitem{lewis2021characterizing}
G.~A. Lewis, S.~Bellomo, and I.~Ozkaya, ``Characterizing and detecting mismatch in machine-learning-enabled systems,'' in \emph{2021 IEEE/ACM 1st Workshop on AI Engineering-Software Engineering for AI (WAIN)}.\hskip 1em plus 0.5em minus 0.4em\relax IEEE, 2021.

\bibitem{heyn2023investigation}
H.-M. Heyn, E.~Knauss, I.~Malleswaran, and S.~Dinakaran, ``An investigation of challenges encountered when specifying training data and runtime monitors for safety critical ml applications,'' in \emph{International working conference on requirements engineering: foundation for software quality}.\hskip 1em plus 0.5em minus 0.4em\relax Springer, 2023.

\bibitem{schroder2022monitoring}
T.~Schr{\"o}der and M.~Schulz, ``Monitoring machine learning models: a categorization of challenges and methods,'' \emph{Data Science and Management}, vol.~5, no.~3, 2022.

\bibitem{karval2023catching}
R.~Karval and K.~N. Singh, ``Catching silent failures: A machine learning model monitoring and explainability survey,'' in \emph{2023 OITS International Conference on Information Technology (OCIT)}.\hskip 1em plus 0.5em minus 0.4em\relax IEEE, 2023.

\bibitem{studer2021towards}
S.~Studer, T.~B. Bui, C.~Drescher, A.~Hanuschkin, L.~Winkler, S.~Peters, and K.-R. M{\"u}ller, ``Towards crisp-ml (q): a machine learning process model with quality assurance methodology,'' \emph{Machine learning and knowledge extraction}, vol.~3, no.~2, 2021.

\bibitem{kreuzberger2023machine}
D.~Kreuzberger, N.~K{\"u}hl, and S.~Hirschl, ``Machine learning operations (mlops): Overview, definition, and architecture,'' \emph{IEEE access}, vol.~11, 2023.

\bibitem{matsui2022mlops}
B.~M. Matsui and D.~H. Goya, ``Mlops: five steps to guide its effective implementation,'' in \emph{Proceedings of the 1st International Conference on AI Engineering: Software Engineering for AI}, 2022.

\bibitem{zimelewicz2024ml}
E.~Zimelewicz, M.~Kalinowski, D.~Mendez, G.~Giray, A.~P. Santos~Alves, N.~Lavesson, K.~Azevedo, H.~Villamizar, T.~Escovedo, H.~Lopes \emph{et~al.}, ``Ml-enabled systems model deployment and monitoring: Status quo and problems,'' in \emph{International Conference on Software Quality}.\hskip 1em plus 0.5em minus 0.4em\relax Springer, 2024.

\bibitem{ml_monitoring_survey_2025}
\BIBentryALTinterwordspacing
H.~Naveed, ``Ml monitoring practices and challenges survey,'' 2025. [Online]. Available: \url{https://tinyurl.com/bdhmwzmr}
\BIBentrySTDinterwordspacing

\bibitem{qualtrics2025}
{Qualtrics}, ``{Qualtrics XM: The Leading Experience Management Software},'' \url{https://www.qualtrics.com/}, 2025, accessed: 2025-05-19.

\bibitem{prolific2025}
{Prolific}, ``Prolific: Access high-quality data for research and ai training,'' \url{https://www.prolific.com}, 2025, accessed: 2025-05-19.

\bibitem{aws2024}
\BIBentryALTinterwordspacing
{Amazon Web Services}, ``Amazon web services (aws),'' 2024, accessed: 2024-05-23. [Online]. Available: \url{https://aws.amazon.com}
\BIBentrySTDinterwordspacing

\bibitem{openaiapi2024}
\BIBentryALTinterwordspacing
{OpenAI}, ``Openai api,'' 2024, accessed: 2024-05-23. [Online]. Available: \url{https://platform.openai.com}
\BIBentrySTDinterwordspacing

\bibitem{docker2024}
\BIBentryALTinterwordspacing
{Docker Inc.}, ``Docker: Empowering app development for developers,'' 2024, accessed: 2024-05-23. [Online]. Available: \url{https://www.docker.com}
\BIBentrySTDinterwordspacing

\bibitem{kubernetes2024}
\BIBentryALTinterwordspacing
{The Kubernetes Authors}, ``Kubernetes: Production-grade container orchestration,'' 2024, accessed: 2024-05-23. [Online]. Available: \url{https://kubernetes.io}
\BIBentrySTDinterwordspacing

\bibitem{liu2024exploring}
F.~Liu, Y.~Liu, L.~Shi, H.~Huang, R.~Wang, Z.~Yang, L.~Zhang, Z.~Li, and Y.~Ma, ``Exploring and evaluating hallucinations in llm-powered code generation,'' \emph{arXiv preprint arXiv:2404.00971}, 2024.

\bibitem{rezaei2025fairness}
A.~Rezaei~Nasab, M.~Dashti, M.~Shahin, M.~Zahedi, H.~Khalajzadeh, C.~Arora, and P.~Liang, ``Fairness concerns in app reviews: A study on ai-based mobile apps,'' \emph{ACM Transactions on Software Engineering and Methodology}, vol.~34, no.~2, 2025.

\bibitem{mlflow}
\BIBentryALTinterwordspacing
{MLflow Authors}, ``Mlflow: An open source platform for the machine learning lifecycle,'' 2024, accessed: 2024-05-25. [Online]. Available: \url{https://mlflow.org/}
\BIBentrySTDinterwordspacing

\bibitem{wandb}
\BIBentryALTinterwordspacing
{Weights and Biases, Inc.}, ``Weights and biases: Developer tools for machine learning,'' 2024, accessed: 2024-05-25. [Online]. Available: \url{https://wandb.ai}
\BIBentrySTDinterwordspacing

\bibitem{prometheus}
\BIBentryALTinterwordspacing
{Prometheus Authors}, ``Prometheus: Monitoring system \& time series database,'' 2024, accessed: 2024-05-25. [Online]. Available: \url{https://prometheus.io/}
\BIBentrySTDinterwordspacing

\bibitem{grafana}
\BIBentryALTinterwordspacing
{Grafana Labs}, ``Grafana: The open observability platform,'' 2024, accessed: 2024-05-25. [Online]. Available: \url{https://grafana.com/}
\BIBentrySTDinterwordspacing

\bibitem{berger2014kolmogorov}
V.~W. Berger and Y.~Zhou, ``Kolmogorov--smirnov test: Overview,'' \emph{Wiley statsref: Statistics reference online}, 2014.

\bibitem{menendez1997jensen}
M.~L. Men{\'e}ndez, J.~A. Pardo, L.~Pardo, and M.~d.~C. Pardo, ``The jensen-shannon divergence,'' \emph{Journal of the Franklin Institute}, vol. 334, no.~2, 1997.

\bibitem{evidently}
\BIBentryALTinterwordspacing
{Evidently AI}, ``Evidently: Open-source ml monitoring,'' 2024, accessed: 2024-05-25. [Online]. Available: \url{https://www.evidentlyai.com/}
\BIBentrySTDinterwordspacing

\bibitem{alibidetect}
\BIBentryALTinterwordspacing
{Seldon Technologies}, ``Alibi detect: Outlier, adversarial and drift detection library,'' 2024, accessed: 2024-05-25. [Online]. Available: \url{https://github.com/SeldonIO/alibi-detect}
\BIBentrySTDinterwordspacing

\bibitem{azureappinsights}
\BIBentryALTinterwordspacing
{Microsoft Corporation}, ``Azure application insights,'' 2024, accessed: 2024-05-25. [Online]. Available: \url{https://learn.microsoft.com/en-us/azure/azure-monitor/app/app-insights-overview}
\BIBentrySTDinterwordspacing

\bibitem{datadog}
\BIBentryALTinterwordspacing
{Datadog, Inc.}, ``Datadog: Cloud monitoring as a service,'' 2024, accessed: 2024-05-25. [Online]. Available: \url{https://www.datadoghq.com/}
\BIBentrySTDinterwordspacing

\bibitem{tensorrt}
\BIBentryALTinterwordspacing
{NVIDIA Corporation}, ``Nvidia tensorrt: High performance deep learning inference optimizer and runtime,'' 2024, accessed: 2024-05-25. [Online]. Available: \url{https://developer.nvidia.com/tensorrt}
\BIBentrySTDinterwordspacing

\bibitem{onnx}
\BIBentryALTinterwordspacing
{ONNX Community}, ``Onnx: Open neural network exchange,'' 2024, accessed: 2024-05-25. [Online]. Available: \url{https://onnx.ai/}
\BIBentrySTDinterwordspacing

\bibitem{redis}
\BIBentryALTinterwordspacing
{Redis Ltd.}, ``Redis: In-memory data structure store,'' 2024, accessed: 2024-05-25. [Online]. Available: \url{https://redis.io/}
\BIBentrySTDinterwordspacing

\bibitem{pingdom}
\BIBentryALTinterwordspacing
{Pingdom}, ``Pingdom website monitoring,'' 2024, accessed: 2024-05-25. [Online]. Available: \url{https://www.pingdom.com/}
\BIBentrySTDinterwordspacing

\bibitem{awswaf}
\BIBentryALTinterwordspacing
{Amazon Web Services}, ``Aws waf: Web application firewall,'' 2024, accessed: 2024-05-25. [Online]. Available: \url{https://aws.amazon.com/waf/}
\BIBentrySTDinterwordspacing

\bibitem{fairlearn}
\BIBentryALTinterwordspacing
{Fairlearn Contributors}, ``Fairlearn: Assess and improve fairness in ml models,'' 2024, accessed: 2024-05-25. [Online]. Available: \url{https://fairlearn.org/}
\BIBentrySTDinterwordspacing

\bibitem{aif360}
\BIBentryALTinterwordspacing
{IBM Research}, ``Ai fairness 360 (aif360),'' 2024, accessed: 2024-05-25. [Online]. Available: \url{https://research.ibm.com/blog/ai-fairness-360}
\BIBentrySTDinterwordspacing

\bibitem{openaidashboard}
\BIBentryALTinterwordspacing
{OpenAI}, ``Openai api dashboard,'' 2024, accessed: 2024-05-25. [Online]. Available: \url{https://platform.openai.com/account/usage}
\BIBentrySTDinterwordspacing

\bibitem{prinz2021low}
N.~Prinz, C.~Rentrop, and M.~Huber, ``Low-code development platforms-a literature review.'' in \emph{AMCIS}, 2021.

\bibitem{moskal2021no}
M.~Moskal, ``No-code application development on the example of logotec app studio platform,'' \emph{Informatyka, Automatyka, Pomiary w Gospodarce i Ochronie {\'S}rodowiska}, vol.~11, no.~1, 2021.

\bibitem{kourouklidis2020towards}
P.~Kourouklidis, D.~Kolovos, N.~Matragkas, and J.~Noppen, ``Towards a low-code solution for monitoring machine learning model performance,'' in \emph{Proceedings of the 23rd ACM/IEEE international conference on model driven engineering languages and systems: companion proceedings}, 2020.

\bibitem{deepeval}
\BIBentryALTinterwordspacing
{Confident AI}, ``Deepeval: The open-source llm evaluation framework,'' 2024, accessed: 2024-05-25. [Online]. Available: \url{https://github.com/confident-ai/deepeval}
\BIBentrySTDinterwordspacing

\bibitem{gu2024survey}
J.~Gu, X.~Jiang, Z.~Shi, H.~Tan, X.~Zhai, C.~Xu, W.~Li, Y.~Shen, S.~Ma, H.~Liu \emph{et~al.}, ``A survey on llm-as-a-judge,'' \emph{arXiv preprint arXiv:2411.15594}, 2024.

\bibitem{eck2022monitoring}
B.~Eck, D.~Kabakci-Zorlu, Y.~Chen, F.~Savard, and X.~Bao, ``A monitoring framework for deployed machine learning models with supply chain examples,'' in \emph{2022 IEEE International Conference on Big Data (Big Data)}.\hskip 1em plus 0.5em minus 0.4em\relax IEEE, 2022.

\bibitem{sopan2021ai}
A.~Sopan and K.~Berlin, ``Ai total: Analyzing security ml models with imperfect data in production,'' in \emph{2021 IEEE Symposium on Visualization for Cyber Security (VizSec)}.\hskip 1em plus 0.5em minus 0.4em\relax IEEE, 2021.

\bibitem{nogueira2024insights}
V.~L. Nogueira, F.~S. Felizardo, A.~M. Amaral, W.~K. Assun{\c{c}}{\~a}o, and T.~E. Colanzi, ``Insights on microservice architecture through the eyes of industry practitioners,'' in \emph{2024 IEEE International Conference on Software Maintenance and Evolution (ICSME)}.\hskip 1em plus 0.5em minus 0.4em\relax IEEE, 2024.

\bibitem{peruma2024developer}
A.~Peruma, T.~Huo, A.~C. Ara{\'u}jo, J.~Imanmka, and R.~Kazman, ``A developer-centric study exploring mobile application security practices and challenges,'' in \emph{2024 IEEE International Conference on Software Maintenance and Evolution (ICSME)}.\hskip 1em plus 0.5em minus 0.4em\relax IEEE, 2024.

\end{thebibliography}

\end{document}